\documentclass[
  reprint,
 superscriptaddress,
bibnotes,
 amsmath,amssymb,
 aps,
pra,
]{revtex4-2}
\usepackage{orcidlink}
\usepackage{graphicx}
\usepackage{physics}
\usepackage{xspace}
\newcommand{\ps}{phase space\xspace}
\usepackage{hyperref}

\usepackage{color}
\usepackage{soul}
\sethlcolor{yellow}
\soulregister\cite7
\soulregister\ref7
\soulregister\pageref7
\soulregister\flalign7
\soulregister\xspace7
\soulregister\ps7

\newcommand{\VEC}[1]{{\mbox{\boldmath${#1}$}}}

\usepackage[normalem]{ulem}

\makeatletter
\newsavebox{\@brx}
\newcommand{\llangle}[1][]{\savebox{\@brx}{\(\m@th{#1\langle}\)}%
  \mathopen{\copy\@brx\kern-0.5\wd\@brx\usebox{\@brx}}}
\newcommand{\rrangle}[1][]{\savebox{\@brx}{\(\m@th{#1\rangle}\)}%
  \mathclose{\copy\@brx\kern-0.5\wd\@brx\usebox{\@brx}}}
\makeatother

\usepackage{forloop,ifthen} 

\usepackage{xifthen}
\newcommand{\refAppendix}[6]{#1
  \ifthenelse{\isempty{#2}}%
    {}
    {\protect\cite{#2}}
    #3\protect\ref{#4}#5#6\xspace
}

\def\state{{\hat \rho}}
\def\refl{ {\varrho}}
\def\tr{\operatorname{Tr}}

\newcommand{\gammaSubs}{\frac{\pi}{2}}

\newcommand{\xA}[1]{x_{#1}'}
\newcommand{\pA}[1]{p_{#1}'}

\def\rV{{\VEC{r}}}
\def\rVf{{\VEC{r}'}}

\begin{document}

\title{Wigner's Phase Space Current for Variable Beam Splitters
  \\
  --- Phase Space Rotations and Newtonian Trajectories --- }

\author{Ole Steuernagel\orcidlink{0000-0001-6089-7022}}
\email{Ole.Steuernagel@gmail.com}
\affiliation{Institute of Photonics Technologies, National Tsing Hua University, Hsinchu 30013, Taiwan}

\author{Hsien-Yi Hsieh\orcidlink{0000-0001-5227-8248}}
\affiliation{Institute of Photonics Technologies, National Tsing Hua University, Hsinchu 30013, Taiwan}

\author{Hua-Li Chen} 
\affiliation{Department of Physics, National Tsing Hua University, Hsinchu 30013, Taiwan}

\author{Ray-Kuang Lee\orcidlink{0000-0002-7171-7274}}
\email{rklee@ee.nthu.edu.tw}
\affiliation{Institute of Photonics Technologies, National Tsing Hua University, Hsinchu 30013, Taiwan}
\affiliation{Department of Physics, National Tsing Hua University, Hsinchu 30013, Taiwan}
\affiliation{Center for Theory and Computation, National Tsing Hua University, Hsinchu 30013, Taiwan}
\affiliation{Center for Quantum Science Technology, Hsinchu 30013, Taiwan}
 
\date{\today}
\begin{abstract}
  Beam splitters allow us to superpose two continu\-ous single mode quantum systems. To study the
  behaviour of beam splitters' strongly mode mixing dynamics we consider variable beam splitters
  acting on Wigner's \ps distribution,~$W$, the evolution of which is governed by the
  continuity-equation $ \frac{\partial}{\partial \tau} W = - \VEC{\nabla } \cdot \VEC{J}$. We derive
  the form of the corresponding Wigner current,~$\VEC{J}$. $\VEC{J}$'s form allows us to use a
  classical trajectories-approach to analyze the influence of the two modes on each other. We show
  that the dynamics for variable beam splitters amounts to a rotation confined within the plane of
  the two positions together with the same simultaneous rotation confined within the plane of the
  two momenta. In this way explicit and very transparent expressions for the rotated Wigner
  distributions and Wigner currents can be given in terms of classical trajectories. This helps us
  to gain deeper insights and perform geometrical analyses of the mixing of modes at beam splitters.
\end{abstract}

\maketitle

\section{Introduction}

In quantum optics beam splitters are often used to entang\-le two matched
modes~\cite{Titulaer_Glauber__PR66,HongOuMandel87}. Such studies traditionally focus on the states
alone, typically in Fock-space or using \ps
representations~\cite{Campos_PRA89,Leonhardt_PRA93,Dakna__EPJD98}.

Here, we show that, instead of focusing on the state alone, it can be useful to complement such
studies by describing and visualizing the dynamics of the beam splitter interaction using Wigner's
\ps
current,~$\VEC{J}$,~\cite{Bauke_2011arXiv1101.2683B,Ole_PRL13,Oliva_Kerr_18,Braasch_PRA19,Chen_PRA23},
thus presenting beam splitter behavior in a new light.

Wigner's description of quantum
systems~\cite{Wigner_PR32,Hillery_PR84,Hirshfeld_AJP02,Hancock_EJP04,Case_AJP08,WignerSpecialfootnote_4D_BS_J_current}
in phase space, based on Wigner's distribution $W(x,p;t)$ ($x, p$ and $t$ are position, momentum and
time), has given us visualizations comparing classical with quantum
states~\cite{Berry_JPA79,Korsch_PD81,Leonhardt_PQE95,Schleich_01,Zurek_NAT01,Zachos_book_21}.
Moreover, the associated Wigner current, $\VEC{J}$, can be constructed from a variable beam
splitter's fictitious-time
evolution~\cite{Lvovsky_16squeezed,Chen_PRA23,Ole__JOSAB25_J_BeamSplitter} and allows for a direct
visualization of the system dynamics and its compari\-son with classical Hamiltonian
flows~\cite{Nolte_PT10,Ole_PRL13,Kakofengitis_EPJP17,Oliva_Shear_19}. No such current exists to
describe $\varrho$'s von Neumann time evolution equation,
$ \frac{\partial \hat \varrho}{\partial \tau} = \frac{1}{{\rm i}\hbar} [\hat H, \hat \varrho]$, (no
such geometric meaning has been attached to studies of the commutator~$[\hat H, \hat \varrho]$ by
itself).

The \ps for two optical modes is four-dimensional, we derive the expressions for the asso\-ciated
Moyal bracket~\cite{Hancock_EJP04,Moyal_MPCPS49,Groenewold_Phys46,Zachos_book_21} and the form of
its four-component Wigner \ps current.

Interestingly, it turns out, as we show here, that keeping all four coordinates for both modes,
gives us a mathe\-matical\-ly simpler description which can even be treated using classical
trajectories~\cite{Takabayasi_PTP54,Oliva_PhysA17}, than the single-mode case (after the other mode
is traced out). For such single-mode cases an equally simple classical description does not
exist~\cite{Ole__JOSAB25_J_BeamSplitter,Oliva_PhysA17}.

Classical mechanics has benefited enormously from the study of classical \ps
portraits~\cite{Nolte_PT10}.  In the quantum case, \ps-based approaches allow for the study of
`quantum phase portraits' (collections of momentary snapshots of field lines arising from the
integration of the vector field $\VEC{J}$~\cite{Ole_PRL13,Kakofengitis_EPJP17}). There is no other
formulation of quantum theory~\footnote{There are several different formulations of quantum theory,
  Schr{\"o}dinger's, Heisenberg's, the \ps formulations due to Wigner, Moyal and Groenewold (and
  their offshoots due to Glauber and Sudarshan, Husimi and others), Feynman's path integrals, the
  de~Broglie--Bohm and many-worlds interpretation, spontaneous collapse models and others.  Here we
  only use Wigner's formulation~\cite{Styer__AJP02,WignerSpecialfootnote_4D_BS_J_current}.}  that
allows for such an intuitive way of studying quantum dynamics and is so reminiscent of classical
dynamics studies~\cite{Hancock_EJP04}. In the Conclusions~\ref{sec:Conclude}, we use this to apply
geometrical reasoning to shed new light on the behaviour of beam splitters.

Before this, we introduce some aspects of the Wigner \ps description of quantum dynamics in
Sect.~\ref{sec:WignerGeneral}, its \ps current,~$\VEC{J}$, in Sect.~\ref{sec:J}, and determine the
explicit general solutions in terms of Newtonian trajectories in Sect.~\ref{subsec:NewtonTraj} and
we discuss connections with experiments in Sect.~\ref{sec:measureQuadratures}.

\section{Wigner distribution and its Continuity Equation\label{sec:WignerGeneral}}

{The time-evolution of Wigner's quantum \ps distribution
$W(\rV,\tau)$~\cite{Case_AJP08,Hillery_PR84}, for a two-mode continuous system with coordinate
vectors~$\rV = (x_a, p_a, x_b, p_b)$, is governed by its \ps
current~$\VEC{J}$ and obeys the continuity equation~\cite{Oliva_PhysA17,Donoso_PRL01,Skodje_PRA89}}
\begin{eqnarray}\label{eq:continuity}
  \frac{\partial W(\rV,\tau)}{\partial \tau} + \VEC{\nabla}
  \cdot \VEC{J}(\rV,\tau) = 0 \; .
\end{eqnarray}
Using the notation $\frac{\partial}{\partial x} = {\partial_{x}}$,
$\VEC{\nabla} = ({\partial x_a}, {\partial p_a}, {\partial x_b}, {\partial p_b})$ is the \ps
divergence operator with respect to modes~$a$ and~$b$, $\tau$ is time,
and~$\VEC{J}(\rV,\tau)=(J_{x_a},J_{p_a},J_{x_b},J_{p_b})^{\bf \intercal}$ is a functional of $W(\rV,\tau)$ and the
system hamiltonian~$H(\rV,\tau)$.

Sufficiently small varia\-tions in reflectivity $|\refl |^2$ allow us to approximate a beam
splitter's mode mixing with a continuous description using an effective
hamiltonian~\cite{Chen_PRA23}, such as expression~(\ref{eq:H_M}) below, and thus continuity
equation~(\ref{eq:continuity})~\cite{Ole__JOSAB25_J_BeamSplitter}.

\subsection{Moyal's bracket~\label{sec:Moyal-bracket}}

We will now remind the reader of how Wigner's and Schr{\"o}dinger's formulation of quantum theory
are connected mathematically~\cite{Zachos_book_21}.

Consider a single-mode operator given in coordinate
representation~$\langle x-y| \hat O | x+y \rangle = O(x-y,x+y)$. To map to Wigner's \ps formulation
we employ the
Wigner-transform,~${\cal W}[\hat O]$,~\cite{Hancock_EJP04,Cohen_LectureNotes18,Zachos_book_21}
\begin{align}\label{eq:WignerWeyl_Trafo}
  {\cal W}[\hat O](x,p) = \int_{-\infty}^\infty dy\; O(x-\frac{y}{2},x+\frac{y}{2})\; {\rm e}^{\frac{{\rm i}}{\hbar} p y}\, .
\end{align}

If $\hat O$ is a (normalized) single-mode density matrix~$\hat \varrho$, then the associated normalized
distribution in the Wigner formulation is the Wigner distribution
\begin{align}\label{eq:WignerDistribution}
  W(x,p,\tau) \equiv  \frac{ {\cal W}[\hat \varrho] }{2 \pi \hbar} \; .
\end{align}

Assuming that the hamiltonian ${\cal W}[ \hat H(\hat x, \hat p) ] = H(x,p)$ is smooth enough,
namely, has a global Taylor expansion, the Wigner transform of the von~Neumann time evolution
equation
\begin{equation}\label{eq:W_of_vNeumann}
  {\cal W}\left[ \frac{\partial \hat \rho}{\partial {t}} = \frac{1}{{\rm i}\hbar} [\hat H, \hat
    \rho] \right]
  \end{equation}
is the Wigner-transform of the von~Neumann equation
\begin{equation}\label{eq:moyal_motion}
  \frac{\partial W}{\partial {t}} = \{\!\!\{ {H} , W \}\!\!\}  = \frac{1}{\rm i \hbar} \left( H \star W - W \star H\right)\; ,
\end{equation}
where the Groenewold-$\star$-product~\cite{Groenewold_Phys46} is the \emph{Wigner transform of operator
  concatenation} translating the non-commutativity of operator multiplication into quantum \ps
${\cal W}\left[ \hat f \circ \hat g \right] = f \star g $.  `$\star$' is given
by~\cite{Groenewold_Phys46,Hirshfeld_AJP02,Zachos_book_21}
\begin{align}\label{Eq:GroenewoldStar}
  \star &\equiv \exp\left[\frac{i\hbar}{2} \overleftrightarrow {\partial} \right]
          = \sum_{n=0}^{\infty} \frac{(i\hbar \overleftrightarrow {\partial})^n}{2^n n!} \; .
\end{align}
Here
$\overleftrightarrow{\partial} = \!\!( \overleftarrow{\partial_x} \overrightarrow{\partial_p} -
\overleftarrow{\partial_p} \overrightarrow{\partial_x} )\!$ is the differential operator in the
Poisson bracket of classical mechanics: $\{f,g\} = f \overleftrightarrow{\partial} g$, and we use
the shorthand~$\frac{\partial}{\partial_z} = \partial_z$, with arrows indicating the `directions' of
differentiation:
$f\overrightarrow{\frac{\partial}{\partial x}} g = g\overleftarrow{\frac{\partial}{\partial x}} f =
f \frac{\partial}{\partial x} g$.

In other words, the non-commutative nature of the composition of Hilbert space operators manifests
itself in the non-commutative structure of Groenewold's
$\star$-pro\-duct~(\ref{Eq:GroenewoldStar}) in the Wigner formulation~\cite{Zachos_book_21}. The
$\star$-pro\-duct is the key-ingredient to map quantum behaviour into
\ps~\cite{Hirshfeld_AJP02,Hancock_EJP04,Zachos_book_21}.

Eq.~(\ref{eq:moyal_motion}) also defines Moyal's
bracket,~$\{\!\!\{ , \}\!\!\}$,~\cite{Moyal_MPCPS49,Zachos_book_21}, which is the quantum version of
Poisson's bracket,~$\{ , \}$, of classical mechanics.

If $f$ or $g$ are polynomials of at most quadratic order in $x$ and $p$ then Moyal's bracket equals
Poisson's bracket. If a hamiltonian is of such form then the quantum dynamics is of classical
form~\cite{Takabayasi_PTP54,Oliva_PhysA17,Ole_FreeHOSC_14} and trajectories exist~\cite{Oliva_PhysA17}.

For a two-mode system, with
$\rV = (x_a, p_a, x_b, p_b)^{\bf \intercal}$, Moyal's bracket,~$\{\!\!\{ , \}\!\!\}$ has the product
form~\cite{Polkovnikov__AnnP10}
\begin{align}\label{EqMoyalBraket4D}
  \!\!  \{\!\!\{ f, g\}\!\!\}  = f(\rV)  \frac{2}{\hbar}
  & \sin\!\!\left[\! \frac{\hbar}{2} \!\! \left( 
    \overleftarrow{\frac{\partial}{\partial {x_a}}} \overrightarrow{\frac{\partial}{\partial p_a}}
    - \overleftarrow{\frac{\partial}{\partial p_a}} \overrightarrow{\frac{\partial}{\partial x_a}}
    \right)\!\!\right]
    \nonumber \\
  \times & \sin\!\!\left[\! \frac{\hbar}{2} \!\!\left( 
           \overleftarrow{\frac{\partial}{\partial x_b}} \overrightarrow{\frac{\partial}{\partial p_b}}
           - \overleftarrow{\frac{\partial}{\partial p_b}} \overrightarrow{\frac{\partial}{\partial x_b}}
           \right) \!\!  \right] g(\rV) ,
\end{align}
which represents the mapping of the tensor product of the Hilbert space underlying a two mode system
to \ps.

\subsection{Beam Splitter and its effective Hamiltonian}

Consider an ensemble of measurements performed on a system of two bosonic modes~$\hat a$
and~$\hat b$ obeying the standard commutation relations $[\hat a, \hat a^\dagger] = 1$ and
$[\hat b^\dagger, \hat b] = -1$, which, with
\begin{flalign}\label{eq:x_p_operators}
  \text{operators for position } \quad \hat x_a & = \tfrac{1}{\sqrt{2}}( \hat a^\dagger + \hat a) \notag \\
  \text{and momentum } \quad \hat p_b & = \tfrac{\rm i}{\sqrt{2}}( \hat b^\dagger - \hat b),
\end{flalign}
implies $[\hat p_b, \hat x_b] = -{\rm i}$. We follow standard convention using rescaled units such
that $\hbar =1$, for a more detailed description of connections with electromagnetic field
quantities see~\cite{Lvovsky_Raymer__RMP09}.

Assuming that we have incoming bosonic modes~$\hat a_{\rm in}$ and $\hat b_{\rm in}$ which are
perfectly mode-matched and being mixed through interaction at a `perfect' lossless two-mode beam
splitter with variable transmissivity, we can describe the associated unitary mode mixing
operator~\cite{Campos_PRA89,Leonhardt_RPP03}, transforming the bosonic optical-mode field operators,
as
\begin{align}
  \left(\!\! \begin{array}{c}
          \hat a
          \\
          \hat b 
        \end{array}\!\right)
  = \hat B(\theta) \; 
  \left(\begin{array}{c}
          \hat a_{\rm in} 
          \\
          \hat b_{\rm in}
        \end{array}\!\right) \; \hat B^\dagger (\theta) \; .
 \label{eq:U_on_modes}
\end{align}
Here~$\hat B$, expressed in terms of the mode mixing angle,
$\theta$, is given by~\cite{Campos_PRA89,Leonhardt_PRA93}
\begin{equation}\label{eq:U_M}
  \hat B (\theta) = \exp[ \frac{\theta}{2} ( \hat a_{\rm in} \hat b_{\rm in}^\dagger - \hat a_{\rm in}^\dagger \hat b_{\rm in} )]
  \equiv \exp[ - {\rm i} \, \tau \, \hat H_{M} ] ,
\end{equation}
where we chose the effective hamiltonian as
\begin{align}\label{eq:H_M}
  \hat H_{M} ={\rm i} \frac{\pi}{2} \left( \hat a_{\rm in} \hat b_{\rm in}^\dagger - \hat a_{\rm in}^\dagger \hat b_{\rm in} \right)
  = \frac{\pi}{2} \left( \hat x_{a} \hat p_{b} -\hat p_{a} \hat x_{b}  \right) \; .
\end{align}
This choice [together with the fact that the reflection amplitude
is~$\refl = \sin( \frac{\pi}{2}{\tau})$, and the transmission amplitude
is~$t = \cos(\frac{\pi}{2}{\tau})$] allow us to choose the convenient range-parameterization, for
the fictitious time $\tau \in [0,1]$, such that the mode-mixing behavior ranges from no mixing,
at~$\tau=0$ (complete transparency), via all intermediate values, where the modes are being mixed,
to eventually no mixing, at~$\tau=1$, due to total reflection at the beam splitter.

In an implementation of the variable beam splitter as a two-mode fiber coupler, $\tau$ parametrizes
the coupling length, implemented as a Mach-Zehnder interferometer, it parametrizes changes of the
phase shifter angle~\cite{Reck_Zeiling__PRL94}.

The Wigner transform of the associated von Neumann equation, Moyal's Eq.~(\ref{eq:moyal_motion}),
gives the fictitious-time evolution equation of the two-mode Wigner
distribution~$W = W(x_a,p_a,x_b,p_b,\tau)$ as
\begin{align}\label{eq:vNeumann_M_TrB_0}
  \frac{\partial {\cal W}[\hat \varrho]}{\partial \tau} = {2\pi\hbar}
  \frac{\partial W}{\partial \tau} = {\cal W}\left[ \frac{\pi}{2 {\rm i}\hbar}
  [ \left( \hat x_{a} \hat p_{b} - \hat p_{a} \hat x_{b}  \right), \hat \varrho_{ab}] \right].
\end{align}

\section{Wigner  Current ${\VEC J}$ for Two Modes\label{sec:J}}

Note, because the hamiltonian is bilinear in creation and annihilation
operators~(\ref{eq:U_on_modes}), no `non-classical' terms, depending on $\hbar$ are present
in~(\ref{eq:vNeumann_M_TrB_0})~\cite{Takabayasi_PTP54}, Moyal's bracket equals Poisson's
bracket. This paves the way for a description in terms of Newton trajectories~\cite{Oliva_PhysA17}.

Applying Eq.~(\ref{EqMoyalBraket4D}) leads to the continuity equation
\begin{flalign}
  \label{eq:Moyal_M_TrA_}
  \frac{\partial W }{\partial \tau} = \; \frac{\pi}{2} & \Big[ x_a p_{b} \left\{
    \overleftarrow{\frac{\partial}{\partial x_a}} \overrightarrow{\frac{\partial}{\partial p_a}} -
    \overleftarrow{\frac{\partial}{\partial p_b}} \overrightarrow{\frac{\partial}{\partial x_b}}
  \right\} W \Big. \notag
  \\
  &- \Big. \! p_{a} x_b \left\{ \overleftarrow{\frac{\partial}{\partial x_b}}
    \overrightarrow{\frac{\partial}{\partial p_b}} - \overleftarrow{\frac{\partial}{\partial p_a}}
    \overrightarrow{\frac{\partial}{\partial x_a}} \right\} W \Big] \; ,
\end{flalign}
which can be rewritten as the divergence of Wigner's \ps
current~\cite{Oliva_Kerr_18,Oliva_PhysA17,Ole__AIPA25}, yielding
a continuity equation of the form~(\ref{eq:continuity})
\begin{flalign}
  \label{eq:Moyal_M_TrB_continuity}
  \frac{\partial W}{\partial \tau} = \frac{\pi}{2} \! \left(\!
    \begin{array}{c} + x_b \\ + p_b \\ -x_a \\ -p_a
    \end{array}\!\right) \! \cdot \! \VEC{\nabla} W = - \VEC{v}  \cdot \! \VEC{\nabla} W = - \VEC{\nabla}  \cdot 
   \VEC{J} \; .
\end{flalign}
Here,~$\VEC{v} = \frac{\pi}{2} (- x_b ,- p_b , x_a, p_a)^{\bf \intercal}$ is the Hamiltonian phase
space velocity vector. In the case of beam splitters it obeys classical
dynamics~\cite{Oliva_PhysA17} and therefore has zero divergence:
$\VEC{\nabla} \!\! \cdot \! \VEC{v} = \VEC{0}$; mani\-festly demonstrating Liouvillian phase space
volume conservation~\cite{Nolte_PT10,Oliva_PhysA17}.

Because $\VEC{\nabla} \!\! \cdot \! \VEC{v} = \VEC{0}$ we can pull $\VEC{v}$ inside, defining the Wigner current as
\begin{flalign}
  \label{eq:_J_B_explicit}
  \VEC{J} = \frac{\pi}{2} \left(\begin{array}{c} - x_b \; W(\rV,\tau) \\ - p_b \;
      W(\rV,\tau) \\ + x_a \; W(\rV,\tau) \\ + p_a \;
      W(\rV,\tau) \end{array}\right) \; .
\end{flalign}

\subsection{Newtonian Trajectories  for Two Modes\label{subsec:NewtonTraj}}

With the initial Wigner distribution, $ W(\rV;0) $ $= W(x_a, p_a, x_b, p_b;0)$, the explicit
solution of Eq.~(\ref{eq:Moyal_M_TrB_continuity}) is
\begin{flalign}
    \label{eq:_W_solution_explicit}
    W(\rVf;\tau) & = W( \rVf(\rV, \tau);0) ,
\end{flalign}
where the functions for coordinates after the beam splitter, $\rVf(\rV,\tau)$, expressed in terms of
the coordinates before the beam splitter, obey
\begin{flalign}
    \label{eq:_R_ps_explicit}
\!\! \rVf(\rV,\tau)
       = \left(\!
         \begin{array}{c}  \xA{a}
           (\tau) \\ {\pA{a}} ( \tau) \\ {\xA{b}} 
      ( \tau) \\ {\pA{b}}  ( \tau)
    \end{array}\!\right)
         = \left(\!
    \begin{array}{c} {x_a} \cos (\gammaSubs \tau) + {x_b} \sin (\gammaSubs \tau) \\ {p_a} \cos (\gammaSubs \tau)+{p_b} \sin
      (\gammaSubs \tau) \\ {x_b} \cos
      (\gammaSubs \tau)-{x_a} \sin (\gammaSubs \tau) \\ {p_b} \cos (\gammaSubs \tau)-{p_a} \sin (\gammaSubs \tau)
    \end{array}\!\right) .
\end{flalign}

We emphasize that this result is general, no special form for the initial state $W(\rV;0)$ (such as
being of unentangled product state-form) has been imposed.

Note that, because of the special form of~$\VEC J$~(\ref{eq:_J_B_explicit}) for beam splitters, the
functions $\xA{a}(\tau)$, $\pA{a}(\tau)$, $\xA{b}(\tau)$ and $\pA{b}(\tau)$ do not mix all
coordinates $(x_a, x_b, p_a, p_b,\tau)$. Instead, in Eqs.~(\ref{eq:_R_ps_explicit}), $\xA{a}$
and~$\xA{b}$ depend on $(x_a, x_b, \tau)$ only, whereas $\pA{a}$ and~$\pA{b}$ depend on
$(p_a, p_b, \tau)$ only.

The inverse to Eq.~(\ref{eq:_W_solution_explicit})
\begin{flalign}
    \label{eq:_W_solution_explicit_inverse}
    W( \rV(\rVf, \tau);\tau) = W(\rV;0) ,
\end{flalign}
uses functions for coordinates {before} the beam splitter, $\rV(\rVf,\tau)$, expressed in terms
of the coordinates after the beam splitter, obeying
\begin{flalign}
    \label{eq:_R_ps_explicit_inverse}
\!\! \rV(\rVf,\tau)
     \!  = \! \left(\!
    \begin{array}{c} {x_a} (\tau) \\ {p_a} ( \tau) \\ {x_b} 
      ( \tau) \\ {p_b}  ( \tau)
    \end{array}\!\right)
         \!\!  = \!\! \left(\!
           \begin{array}{c} {\xA{a}} \cos (\gammaSubs \tau) - {\xA{b}} \sin (\gammaSubs \tau) \\
             {\pA{a}} \cos (\gammaSubs \tau) - {\pA{b}} \sin (\gammaSubs \tau) \\ {\xA{b}} \cos
             (\gammaSubs \tau) + {\xA{a}} \sin (\gammaSubs \tau) \\ {\pA{b}} \cos (\gammaSubs \tau) +
             {\pA{a}} \sin (\gammaSubs \tau)
    \end{array}\!\right)\! .
\end{flalign}

The fact that Eq.~(\ref{eq:_R_ps_explicit}) and~(\ref{eq:_R_ps_explicit_inverse}) are inverse of
each other can be checked by direct substitution: $\rVf(\rV(\rVf,\tau),\tau) = \rVf$ or
$\rV(\rVf(\rV,\tau),\tau) = \rV$.

For the case of a balanced beam splitter, the mode transformations are governed by the coordinate
transformations~(\ref{eq:_R_ps_explicit}) and~(\ref{eq:_R_ps_explicit_inverse}) with
$\tau = \frac{1}{2}$, namely,
\begin{subequations}
  \label{eq:_R_ps_explicit_5050}
\begin{flalign}
  \label{eq:_R_ps_explicit_5050_forward}
  &\rVf(\rV, \tfrac{1}{2}) = \frac{1}{\sqrt{2}} \left(\!
    \begin{array}{c} {x_a} + {x_b}
      \\ {p_a} + {p_b} \\ {x_b}  - {x_a}  \\ {p_b}  - {p_a} 
    \end{array}\!\right)
  \\
  \text{ and \quad} &\rV(\rVf, \tfrac{1}{2}) = \frac{1}{\sqrt{2}} \left(\!
    \begin{array}{c} {x_a'} - {x_b'}
      \\ {p_a'} - {p_b'} \\ {x_b'}  + {x_a'}  \\ {p_b'}  + {p_a'} 
    \end{array}\!\right) \; .
\end{flalign}  
\label{eq:_R_ps_explicit_5050_backward}
\end{subequations}

\section{Connections with measured Quadratures\label{sec:measureQuadratures}}

In balanced homodyne detection (BHD) experiments, the field quadratures
\begin{flalign} \label{eq:_BHD_X}
  \!\!  \langle \hat x_a( \theta_a ) \rangle = \langle \tfrac{\hat a
    {\rm e}^ {-{\rm i} \theta_a} + \hat a^\dagger {\rm e}^ {{\rm i} \theta_a}}{\sqrt 2} \rangle =
  \langle \hat x \cos(\theta_a ) + \hat p \sin ( \theta_a ) \rangle
\end{flalign}
can be measured~\cite{Gerry_Knight__BookQO,Lvovsky_Raymer__RMP09} by the experimenter choosing the
homodyne laser's reference phase,~$\theta_a$.

This, in turn, allows us to reconstruct the quantum state's Wigner distribution in
\ps~\cite{Lvovsky_Raymer__RMP09}.  For this reconstruction, we formally need the probability
distribution pr$(\theta_a, X_a)$ that for a chosen laser reference phase $\theta_a$ the
measurement of $\hat x_a$ yields the outcome $ X_a $. In the experimental implementation the
values of pr$(\theta_a, X_a)$ can only be determined when the varying values of $X_a$ are
resolved. Practically, the measurement of the BHD outcomes are the most straightforward route
towards achieving such resolution, in short, experimentally we determine pr$(\theta_a, X_a)$ from
\begin{subequations}
  \begin{flalign} \label{eq:_BHD_X_Density_rho}
  \!\!  \langle \hat x_a( \theta_a ) \rangle & = \tr \big[ \state \; (\hat x \cos(\theta_a ) + \hat p
  \sin ( \theta_a )) \big]  \\
  & = \int \big. dX_a\big|_{\theta = \theta_a}  \text{pr} (\theta_a, X_a) \times X_a ,
\end{flalign}
\end{subequations}
where $X_a = X_a (\theta_a)$ is the random variable that represents the measured, fluctuating
values of $ \hat x_a( \theta_a ) $ and which is data-logged as a histogram which for many repeated
measurements over the same quantum ensemble, $\state$, approximates the sought after probability distribution
pr$(\theta_a, X_a)$.

Translating Eq.~(\ref{eq:_BHD_X_Density_rho}) for a general operator, $\hat O_a$, to \ps gives us,
according to Eqs.~(\ref{eq:WignerDistribution}) and~(\ref{Eq:GroenewoldStar}),
\begin{subequations}
\begin{flalign} \label{eq:_BHD_OP_Density_W_1}
  \!\!  \langle \hat O_a \rangle & = \iint
  dx_{\theta_a} dp_{\theta_a} W(x_{\theta_a},p_{\theta_a}) \star O(x_{\theta_a},p_{\theta_a})   \\
  & = \iint dx_{\theta_a} dp_{\theta_a} W(x_{\theta_a},p_{\theta_a}) \times
  O(x_{\theta_a},p_{\theta_a}) , \label{eq:_BHD_OP_Density_W_2}
\end{flalign}
\end{subequations}
where we used the shortened notation $\hat x_{ \theta_a} = \hat x_a( \theta_a ) $ and
$\hat p_{ \theta_a} = \hat x_a( \theta_a + \tfrac{\pi}{2}) $, compare Eq.~(\ref{eq:_BHD_X}).

That in the particular case of mapping the \emph{expectation value} of an operator~$\hat O$ to \ps
the $\star$-product in~(\ref{eq:_BHD_OP_Density_W_1}) becomes the scalar product
in~(\ref{eq:_BHD_OP_Density_W_2}), famously, follows from a straightforward but tedious calculation, involving partial integrations,
and is in accord with the cyclicity of the Tr-operation. For the special case of
$\hat O = \hat x_{\theta_a}$ Eq.~(\ref{eq:_BHD_OP_Density_W_2}) becomes
\begin{subequations}
  \label{eq:_BHD_X_Density_W_PS}
  \begin{flalign} \label{eq:_BHD_X_Density_W_PS_1}
  \!\!  \langle \hat x_{\theta_a} \rangle & = \iint dx_{\theta_a} dp_{\theta_a}
  W(x_{\theta_a},p_{\theta_a}) \times x_{\theta_a} 
  \\
  \label{eq:_BHD_X_Density_W_PS_2}
  & = \int dx_{\theta_a}  \; \text{pr}(x_{\theta_a}) \times x_{\theta_a} , 
\end{flalign}
\end{subequations}
where we have used the `projection property' of $W$~\cite{WignerSpecialfootnote_4D_BS_J_current},
namely: integration over the orthogonal \ps coordinate gives a quantum-distribution's correct
marginals
$\langle x_{\theta_a} | \state | x_{\theta_a}\rangle = \text{pr}(x_{\theta_a}) = \int dp_{\theta_a}
W(x_{\theta_a},p_{\theta_a}) $~\cite{Wigner_PR32,Zachos_book_21,Schleich_01}.

Slightly generalizing these considerations to two modes, we denote the measured \ps quadratures by
$x_{\theta_a}$ and $x_{\theta_b}$, respectively, assuming that the local oscillator phases of the
BHDs' are chosen as $\theta_a$ and $\theta_b$.  To capture the quantum states' correlations,
measurements of $x_{\theta_a}$ and $x_{\theta_b}$ have to be logged as 4D data points, as `4-tuples',
such as $(\theta_a,X_a,\theta_b,X_b)$, or as \ps locations
$\VEC{R}
= (X_{\theta_a} \cos(\theta_a), X_{\theta_a} \sin(\theta_a), X_{\theta_b} \cos(\theta_b),
X_{\theta_b} \sin(\theta_b))^{\bf \intercal}$.  Note, for fixed $\theta_a$ and $\theta_b$, the
associated \ps locations, $\VEC{R}_{\theta_a,\theta_b}$, form a 2D sheet, parameterized by $X_a$ and
$X_b$, which describes the locations onto which, according to Eq.~(\ref{eq:_BHD_X_Density_W_PS_2}),
the marginals of $W$ are projected.

When the associated fields are mixed at a beam splitter, the locations of these data points become
mapped as described by Eq.~(\ref{eq:_R_ps_explicit}).  A data entry at a projection sheet location
${\VEC{R}'} = (X_a', P_a', X_b', P_b')^{\bf \intercal}$ for a measurement after the beam splitter
can, consequently, be expressed in terms of measurements before the beam splitter by using the
inversion relation~(\ref{eq:_R_ps_explicit_inverse}), namely,
\begin{flalign}
  \label{eq:_R_ps_explicit_X}
  \VEC{R} = \rV({{\VEC{R}'}},\tau) \! =  \!\left(\!
    \begin{array}{cccl}
      {X_a'} \cos (\gammaSubs \tau)& - & {X_b'} & \sin (\gammaSubs \tau)
      \\
      {P_a'} \cos (\gammaSubs \tau)& - & {P_b'} & \sin (\gammaSubs \tau)
      \\
      {X_b'} \cos (\gammaSubs \tau)& + & {X_a'} & \sin (\gammaSubs \tau)
      \\
      {P_b'} \cos (\gammaSubs  \tau)& + & {P_a'} & \sin (\gammaSubs \tau)
    \end{array}\!\right) . 
\end{flalign}

As long as the measurement data are logged as 4-tuples, as described above, mappings of the
form~(\ref{eq:_R_ps_explicit_X}) can be applied preserving all measured correlations.  Specifi\-cally,
data from measurements after mixing at a beam splitter can be post-processed, as long as each
4-tuple stays together.

For example, mixing at a balanced beam splitter can, according to
Eq.~(\ref{eq:_R_ps_explicit_5050}), be `undone' by adding and subtracting the output signals of the
BHDs~\cite{Chen_26_2modeSqueeExp}.

If a \ps reconstruction of quantum states is sought or if their features can straightforwardly be
described with respect to \ps (such as their fluctuations in $X_{\theta}$), then our analysis allows
to re-express post-beam splitter quantities in terms of pre-beam splitter quantities and vice
versa~\cite{Chen_26_2modeSqueeExp}. For example, if it is known that the initial state is a product state
$W(\rV;0) = W_a(x_a,p_a) W_b(x_b,p_b)$, then after a beam splitter the investigated entangled state
can still be processed as if it were present in two unentangled modes if the measurement data are
post-processed since we can use the inversion~(\ref{eq:_R_ps_explicit_X})
in expression~(\ref{eq:_W_solution_explicit_inverse}):
\begin{subequations}
\begin{eqnarray}\label{eq:Factorize_W}
  W({\rV}';\tau) \mapsto & W( \rVf({\rV},\tau);\tau) =  W(\rV;0) 
  \\ & \quad =  W_a(x_a,p_a) W_b(x_b,p_b)\; .
\end{eqnarray}
\end{subequations}


\section{Conclusions\label{sec:Conclude}}

We emphasize three core results of our considerations:

\emph{Firstly}, the full 2-mode \ps description of the mixing behaviour of a perfect beam splitter is formally
very simple, it can be described with classical Newtonian trajectories because the system
Hamiltonian is bilinear in the field operators~(\ref{eq:H_M})
~\cite{Takabayasi_PTP54,Oliva_PhysA17}, also see Eqs.~(\ref{eq:Moyal_M_TrA_})-(\ref{eq:_J_B_explicit}).

Since the governing Hamiltonian induces the evolution~(\ref{eq:_R_ps_explicit}) of classical form
this guarantees Liouvillian \ps volume conservation, the evolution being of classical canonical form
and symplectic. Even more specifically, the reversed trajectories~(\ref{eq:_R_ps_explicit}) can be rewritten
as
\begin{flalign}
  \label{eq:_R_ps_explicit_pseudoHOSC}
  \rVf(\rV,\tau)=  \left(\!
    \begin{array}{c} {\xA{a}} {(\rV, \tau)} \\ {\pA{a}} {(\rV, \tau)} \\ {\xA{b}} {(\rV, \tau)} \\
      {\pA{b}} {(\rV, \tau)}
    \end{array}\!\right) = \left(\!
    \begin{array}{c} {\xA{a}} {(\rV, \tau)} \\ {\pA{a}} {(\rV, \tau)} \\ {\xA{a}} {(\rV, \tau -
        \gammaSubs)} \\ {\pA{a}} {(\rV, \tau - \gammaSubs)}
    \end{array}\!\right) \; .
\end{flalign}
Interpreting Eq.~(\ref{eq:_R_ps_explicit_pseudoHOSC}) mnemonically: the reversed trajectory
functions $\xA{a} , \xA{b}$ behave together as if they formed the momentum and position of an
anti-clockwise rota\-ting \mbox{harmon\-ic oscilla\-tor. The same applies to the pair~$\pA{a} ,
\pA{b}$.} These rotations are stiff and locked together. This, \emph{second\-ly}, implies that the
nonclassical character of the full Wigner \ps distribution across both modes,~$W(\rV,\tau)$, cannot
change due to the mixing of its modes at a beam splitter:

This result was formally proven by Wang~\cite{Wang__PRA02} and has, here, been given a very simple
geometrical explanation.

\emph{Third}, the description of the behaviour of each single mode, upon beam mixing, and after its
partner mode has been traced out, is considerably more involved than the description of the entire
two mode system. Such a single mode case can show strongly non-classical evolution, including
singular \ps volume changes~\cite{Oliva_PhysA17}; this is in stark contrast to the Liouvillian \ps
volume conservation observed in Eqs.~(\ref{eq:Moyal_M_TrB_continuity})
and~(\ref{eq:_J_B_explicit}). Also, other effects that can change the nonclassical nature of the
state very substantially, can occur for each individual mode under the influence of its (traced-out)
partner mode, for details see~\cite{Ole__JOSAB25_J_BeamSplitter}.

\section*{Acknowledgements}

This work is partially supported by the National Science and Technology Council of Taiwan (Nos
112-2123-M-007-001, 112-2119-M-008-007, 114-2112-M-007-044-MY3), Office of Naval Research Global,
and the collaborative research program of the Institute for Cosmic Ray Research (ICRR) at the
University of Tokyo.

\bibliography{Ole_Bibliography}


\end{document}